\documentclass[10 pt,aps,prd]{revtex4}
\usepackage{times}
\usepackage{graphicx}
\usepackage{verbatim}
\usepackage[usenames, dvipsnames]{color}
\usepackage[colorlinks]{hyperref}
\usepackage[percent]{overpic}
\usepackage{rotating}

\def\mete{\setbox0=\hbox{$E$}%
         \hbox{$E$\rlap{\kern -0.6em\raise0.08\ht0\hbox{/}}}}
\def\met{\ensuremath{\mete_T}}

\def\pt{\ensuremath{{p_T}}}

\def\ttbar{\ensuremath{\rm t\bar t}}
\def\bbbar{\ensuremath{\rm b\bar b}}

\def\ppbar{\ensuremath{\rm p\bar p}}
\def\pp{\ensuremath{\rm pp}}

\def\zprime{\ensuremath{\rm{Z^\prime}}}
\def\Mzprime{\ensuremath{M_{\rm{Z^\prime}}}}

\def\roots{\ensuremath{\sqrt{s}}}

\begin{document}

\title{Discovery potential for heavy $\ttbar$ resonances in dilepton+jets final states 
in pp collisions at $\roots = 14$~TeV \\
A Snowmass 2013 Whitepaper}

%\begin{flushright}
%FERMILAB-FN-0973-E \\
%\end{flushright}

\author{{
Ia Iashvili,$^{1}$
Supriya Jain,$^{1}$
Avto Kharchilava,$^{1}$
Harrison B. Prosper$^{2}$
}}

\affiliation{\large{\vspace*{0.1in}
$^{1}$State University of New York at Buffalo \\
$^{2}$Florida State University, Tallahassee 
}}

\date{September 25, 2013}%

\begin{abstract}
{{
We examine the prospects for probing heavy top quark-antiquark ($\ttbar$)
resonances at the upgraded LHC in pp collisions at $\roots$~=~14~TeV. 
Heavy $\ttbar$ resonances ($\zprime$ bosons) are predicted by several theories 
that go beyond the standard model.
We consider scenarios 
in which each top quark decays leptonically, either to an electron or a muon, and the
data sets correspond to integrated luminosities of 
$\int\mathcal{L}\rm{dt} = 300~\rm{fb^{-1}}$ and 
$\int\mathcal{L}\rm{dt} = 3000~\rm{fb^{-1}}$. 
We present the expected 5$\sigma$ discovery potential for a $\zprime$ resonance 
 as well as the expected upper limits at 95\% C.L.
on the $\zprime$ production cross section and mass in the absence of a discovery.}}
\end{abstract}

\maketitle
%%\tableofcontents

\vspace*{3mm}

\clearpage

%------------------------
% Introduction
%------------------------
\section{Introduction}
\label{sec:intro}

An important goal of the LHC research programs is to deepen our understanding
of electroweak symmetry breaking. Electroweak symmetry breaking in the standard model (SM) 
is closely associated with the existence of a neutral Higgs boson. Therefore, the discovery of a new boson~\cite{Aad:2012tfa,Chatrchyan:2012ufa} with properties 
consistent with those of the SM Higgs boson is clearly
a monumental development. However, the top quark, by far the heaviest
known fundamental particle, has a mass close to the electroweak scale, which 
suggests that it too may
play a role in electroweak symmetry breaking. This 
 alone provides ample motivation for the continued intense scrutiny of
 the top quark in all of its manifestations.
 
A generic prediction of many models that go beyond the standard model (BSM) 
is  the existence of at least one heavy neutral boson, referred to generically as a
$\zprime$, that preferentially
decays to a $\ttbar$ pair and that appears as a resonant structure superimposed
on the SM  $\ttbar$ continuum production. These models include
coloron models~\cite{laneref31,theory_hill_parke,laneref32,jainharris},  models based on
extended gauge theories with massive color-singlet Z-like
bosons~\cite{theory_rosner,theory_lynch,theory_carena}, and
models in which
a pseudoscalar Higgs boson may couple strongly to top quarks~\cite{theory_dicus}.
Furthermore,  various extensions of the Randall-Sundrum model~\cite{theory_randall_sundrum}
with extra dimensions predict Kaluza-Klein  excitations of
gluons~\cite{theory_agashe}, or gravitons~\cite{theory_davoudiasl}, both of
which can have enhanced couplings to $\ttbar$ pairs.
The recent observation of forward-backward asymmetry in $\ttbar$ production
at the Tevatron~\cite{cdf_ttbar_asymmetry1,d0_ttbar_asymmetry1,
cdf_ttbar_asymmetry2,d0_ttbar_asymmetry2} has inspired new models~\cite{theory_bai,
theory_frampton,theory_gresham,theory_antunano,theory_alvarez} that
explain the observation by positing new physics at the TeV scale. The latter can
manifest itself as a broad enhancement over the SM $\ttbar$ production
at high invariant mass. The top quark, and  $\ttbar$ production in particular, is a powerful
probe of potential new physics.

Direct searches for heavy $\ttbar$ resonances have been performed
at the Tevatron and the LHC. No such resonances have been found.  
The Tevatron experiments probed the mass range
up to $\sim$900 GeV~\cite{cdfZprime,d0Zprime},
while the LHC experiments have set sub-pb limits
on the $\ttbar$ resonance production cross section in the mass range of 1--3 TeV depending on the $\zprime$
width, and have excluded the existence of a narrow width $\zprime$ ($\Gamma_{\zprime} = 0.012M_{\zprime}$) below
$\Mzprime$ = 2.1~TeV at 95\% C.L.\cite{cms_ttbar_resonance1,atlas_ttbar_resonance1, 
atlas_ttbar_resonance2, cms_ttbar_resonance2, cms_ttbar_resonance3, cms_ttbar_resonance4}.
 
The null results indicate that $\ttbar$ resonances, if they exist, must have
masses in the TeV range or higher. In this paper, we  
examine how high a mass can be expected to be probed  using
$\zprime$ $\to$ $\ttbar$ $\to$ W$^+$b W$^-$ $\mathrm{\bar{b}}$ production in
 $\pp$ collisions at the upgraded LHC operating at $\roots$~=~14~TeV. We consider final
states in which both W bosons decay to leptons (electron or muon), that is, final
states  comprising
two high $\pt$ leptons of opposite charge (e$^+$e$^-$, $\mu^+\mu^-$, or
e$^\pm\mu^\mp$), at least two jets from the hadronization of $\mathrm{b/\bar{b}}$ quarks,
and missing transverse momentum due to escaping neutrinos.
Top quarks from the decay of a heavy $\zprime$ are
expected to be highly boosted leading to 
decay products that may not be spatially well separated. Consequently, we expect 
events would contain a non-isolated lepton from W$\to \ell \nu$
decay that is partially or fully overlapped with the b-quark jet from 
t$\to$ Wb decay.

 The dominant
(irreducible) background is the $\ttbar$ continuum production. Other
SM processes contributing to the background are the production of single top quarks,
$\rm{Z}/\gamma^*/\rm{W}$+jets, and dibosons (WW, WZ, and ZZ). 
We consider two potential data sets, one corresponding to
$\int\mathcal{L}\rm{dt} = 300~\rm{fb^{-1}}$ and the other to 
$\int\mathcal{L}\rm{dt} = 3000~\rm{fb^{-1}}$,
as anticipated by the end of Run 2 of the upgraded LHC and by the end of
the High Luminosity LHC (HL-LHC) runs, respectively.

% NEXT SECTION

%----------------------------
% Signal and Background Samples
%----------------------------
\section{Signal and Background Samples}
\label{sec:modeling}

This study considers four different $\zprime$ mass hypotheses,  $\Mzprime$~=~2, 3, 4 and 5~TeV, and assumes a
resonance width of $\Gamma_{\zprime} = 0.012M_{\zprime}$. For each mass hypothesis, 
signal event samples are generated using the {\sc pythia} program~\cite{Pythia8}.
The expected signal yields are computed
using the leading-order (LO) cross sections for a leptophobic $\zprime$~\cite{jainharris} scaled by a K-factor of
1.3~\cite{ZprimeKfactor} to approximate the cross section at next-to-leading-order (NLO).
The SM background samples are generated using the {\sc Madgraph} event generator~\cite{SMbkg} 
 and higher-order and non-perturbative effects are approximated using {\sc pythia}
through its parton showering and hadronization models.
The LO cross sections for the background processes are 
obtained from the event generator and corrected for
NLO effects~\cite{snowmass_mc}.
The detector response to the simulated events is computed using the
``Combined Snowmass LHC detector"~\cite{SnowmassSimulation},
which is implemented in the
{\sc Delphes-3} fast simulation program~\cite{delphes}.
The {\sc Delphes-3} program can be used to model (to an accuracy of about 10 -- 20\%)
the
projected performance of future ATLAS~\cite{atlas} and CMS~\cite{cms} detectors
at the upgraded LHC. The program also supports the simulation of additional pp
interactions per bunch crossing (that is, in-time pile-up). 
We use samples that correspond to two different luminosity and pile-up (PU) scenarios 
at $\roots$~=~14~TeV:
$\int\mathcal{L}\rm{dt} = 300~\rm{fb^{-1}}$, with an average number of 
pile-up events of $<\rm{PU}>$ = 50 events per bunch crossing (LHC Run 2), and 
$\int\mathcal{L}\rm{dt} = 3000~\rm{fb^{-1}}$, with $<\rm{PU}>$ = 140 events per bunch
crossing (HL-LHC). 
% 

% NEXT SECTION

%----------------------------
% Event Selections and Yields
%----------------------------
\section{Event Selection and Yields}
\label{sec:selection}

We select $\zprime \rightarrow \ttbar \rightarrow 2\ell+2\nu+\bbbar$
candidate events by requiring two oppositely charged leptons, each with
$\pt > 20$~GeV and pseudorapidity $|\eta| < 2.4$, and at least
two jets within $|\eta| < 2.4$ and with $\pt > 30$~GeV.
In addition, events are required to have $\met > 30$~GeV and 
at least one b-tagged jet, where the b-tagging efficiency is assumed to be
$\sim 65\%$~\cite{snowmass_mc}.  In order to reduce the background from low-mass dilepton
resonances, events are rejected if the dilepton mass $M_{\ell\ell} < 12$~GeV. 
The remaining events are split into three disjoint categories depending on the
lepton flavors, the $ee$, $\mu\mu$, and $e\mu$ channels. In the $ee$ and $\mu\mu$ channels, the contribution from Z+jets
production is suppressed by vetoing events with $76 < M_{\ell\ell} < 106$~GeV. 
We refer to the sample at this stage as the  ``pre-selected" 
sample.

%The $\zprime$ signal events in the dilepton channel are selected by requiring
%two leptons (electron or muon) of opposite charge within pseudorapidity $|\eta| < 2.4$ 
%with transverse momenta of $\pt > 100$~GeV and $>$~20~GeV. Each event is 
%required to have at least two jets within $|\eta| < 2.4$ and with $\pt > 100$~GeV
%and $>$~50GeV. High $\pt$ requirements on leptons and jets are 
%justified by expected signal kinematics, and at the same
%time is dictated by the expected high thresholds on triggering objects
%at high luminosity LHC environment.cSelected events are spilt
%into three non-overlapping categories depending on the lepton flavors:
%$ee$, $\mu\mu$, and $e\mu$ channels.
%% to enhance the sensitivity due to
%% different lepton-flavor response in the detector.
%%

Starting with the pre-selected sample, selection cuts are optimized
using the Random Grid Search (RGS) method~\cite{rgs} and the signal
significance measure 
$S/\sqrt{B}$, where S is the expected number of $\zprime$ signal events with $\Mzprime = 2$~TeV,
and $B$ is the total expected background. Since the signal-to-background separation power increases
with the hypothesized $\zprime$ mass, the set of cuts optimized for $\Mzprime = 2$~TeV also yields
good discrimination between signal and  background  for higher $\Mzprime$
values.
The selection optimization is performed separately for the
$\int\mathcal{L}\rm{dt} = 300~\rm{fb^{-1}}$ and
$\int\mathcal{L}\rm{dt} = 3000~\rm{fb^{-1}}$ scenarios.

\begin{figure}[hbt]
 \centering
  \includegraphics[width=0.5\textwidth]{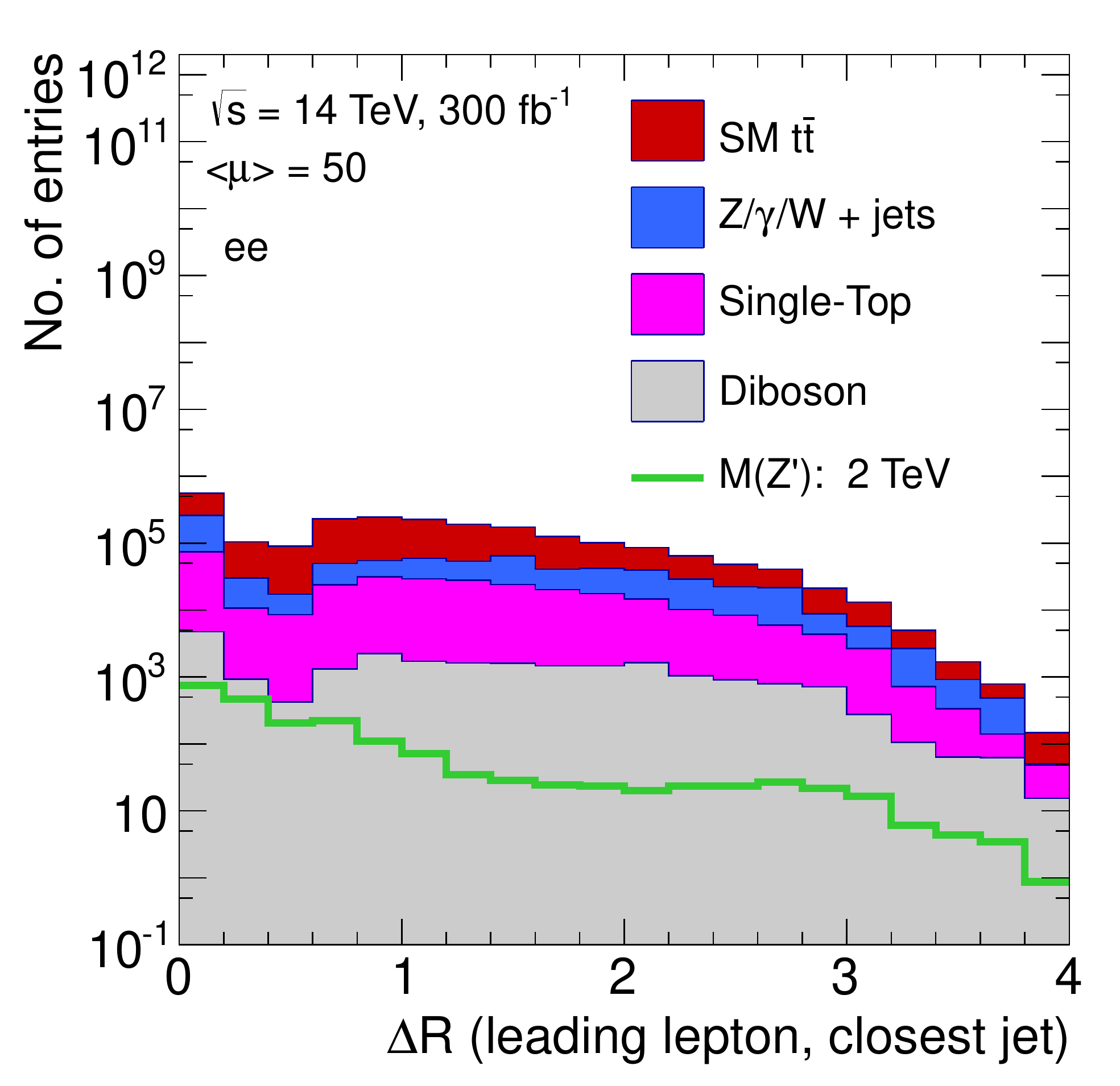}
\vspace{-0.12in}
 \caption{Distribution of $\Delta R$ between the leading lepton
  and closest jet in the $ee$ channel in the pre-selected cuts sample.
  Shown are contributions from the SM background processes and the $\zprime$ signal 
  assuming $\Mzprime$ 2~TeV and
  $\int\mathcal{L}\rm{dt} = 300$~$\rm{fb^{-1}}$ luminosity.
\label{fig:dRJetLep1_b1}}
\end{figure}

The kinematic variables used in the RGS procedure are the 
transverse momenta of the two leading leptons and the two leading jets, and the
missing transverse momentum. In addition,  we use two highly
discriminating variables. The first is the separation between the
lepton and the closest jet in the space $\Delta R=\sqrt{\Delta\eta^2 + \Delta\phi^2}$,
where $\Delta\eta$ and $\Delta\phi$ are the pseudorapidity and the azimuthal angle differences, respectively,
between the lepton and jet. 
The boosted top quarks from the decay of a heavy $\zprime$ produce
a lepton and b-quark that are close together in space. We therefore expect $\Delta R$
to be smaller on average for the signal than for the background processes, which, unlike
the signal, do not contain
highly boosted particles.
Figure~\ref{fig:dRJetLep1_b1} shows an example of the distribution of the $\Delta R$
between the leading lepton and the closest jet in the $ee$
channel in the pre-selected sample.  

\begin{figure}[hbt]
 \centering
  \includegraphics[width=0.33\textwidth]{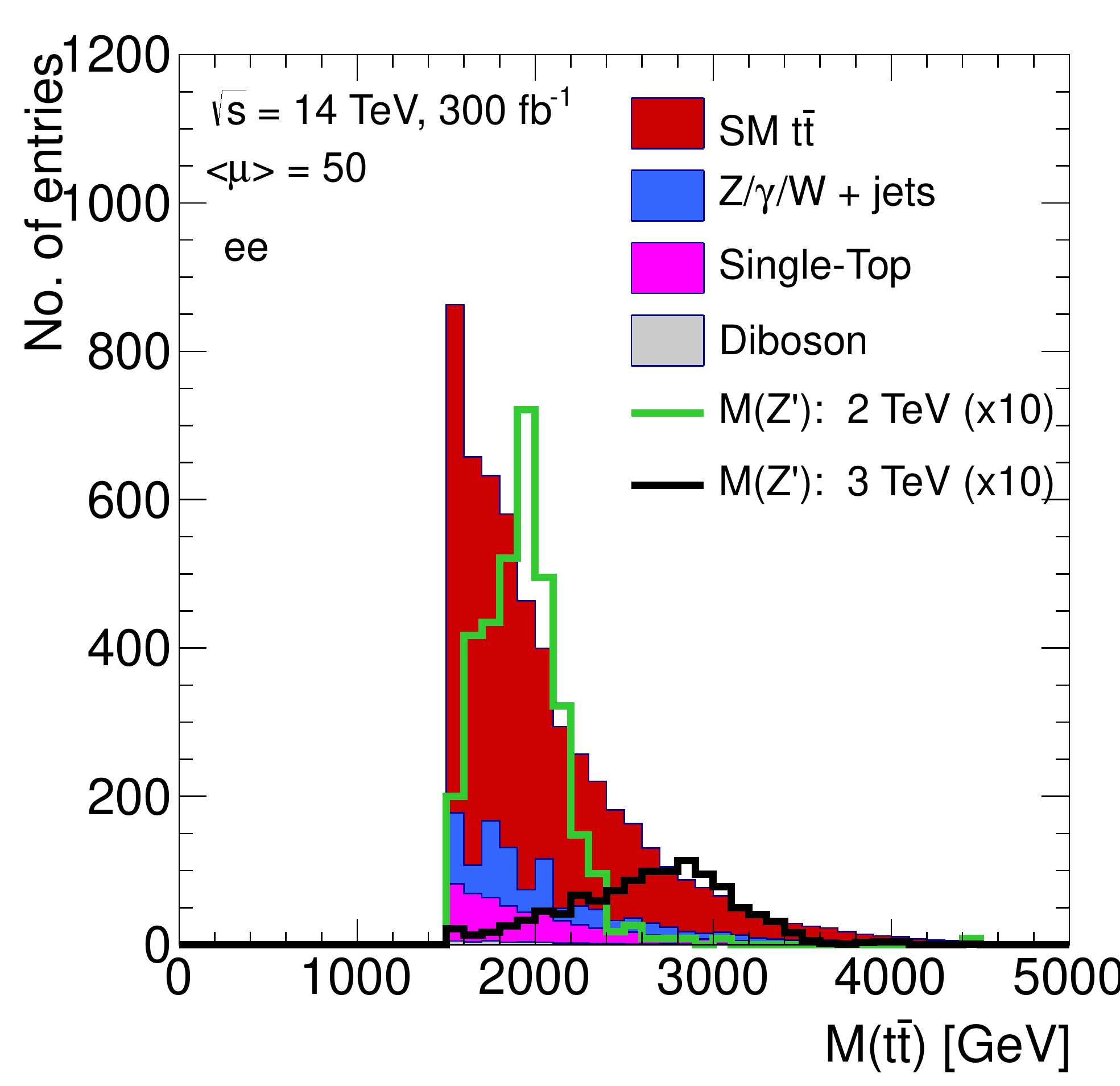}
  \includegraphics[width=0.33\textwidth]{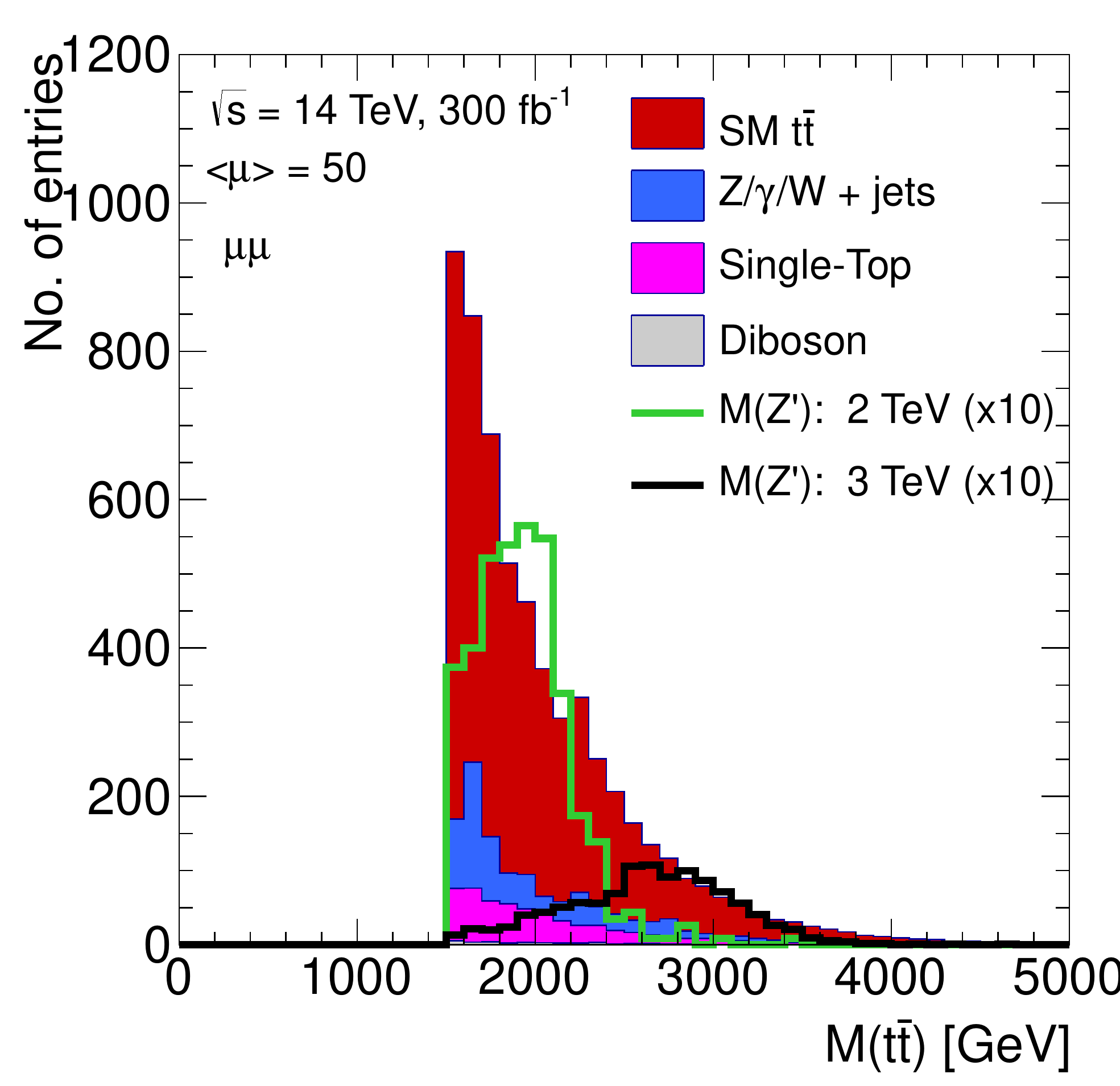}
  \includegraphics[width=0.33\textwidth]{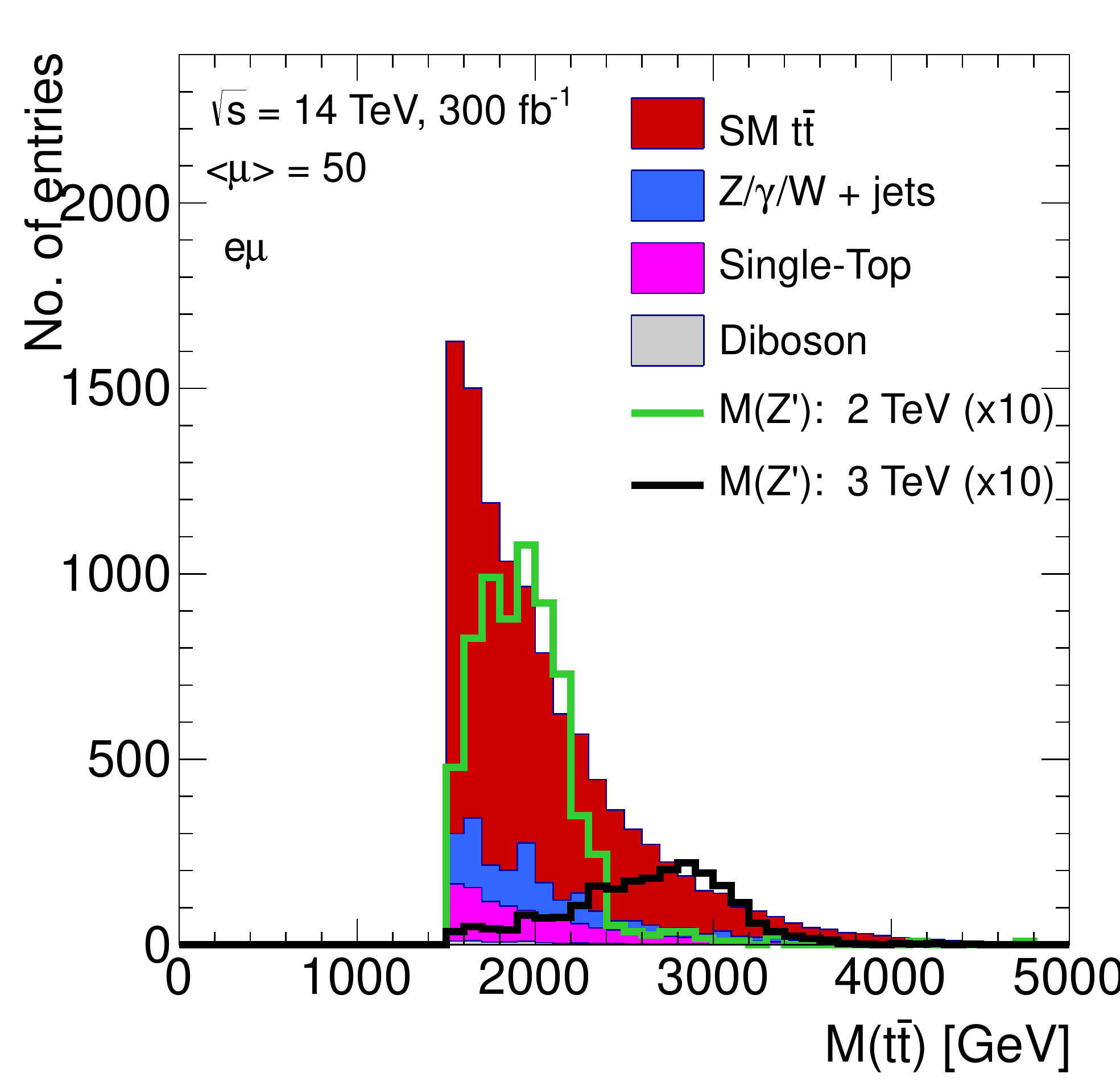}
\vspace{-0.12in}
 \caption{Distributions of the mass variable $M$ for the $ee$,
   $\mu\mu$, and $e\mu$ channels for 300~$\rm{fb^{-1}}$ after selection cuts are applied. 
\label{fig:invmass300}}
\end{figure}

\begin{figure}[hbt]
 \centering
  \includegraphics[width=0.33\textwidth]{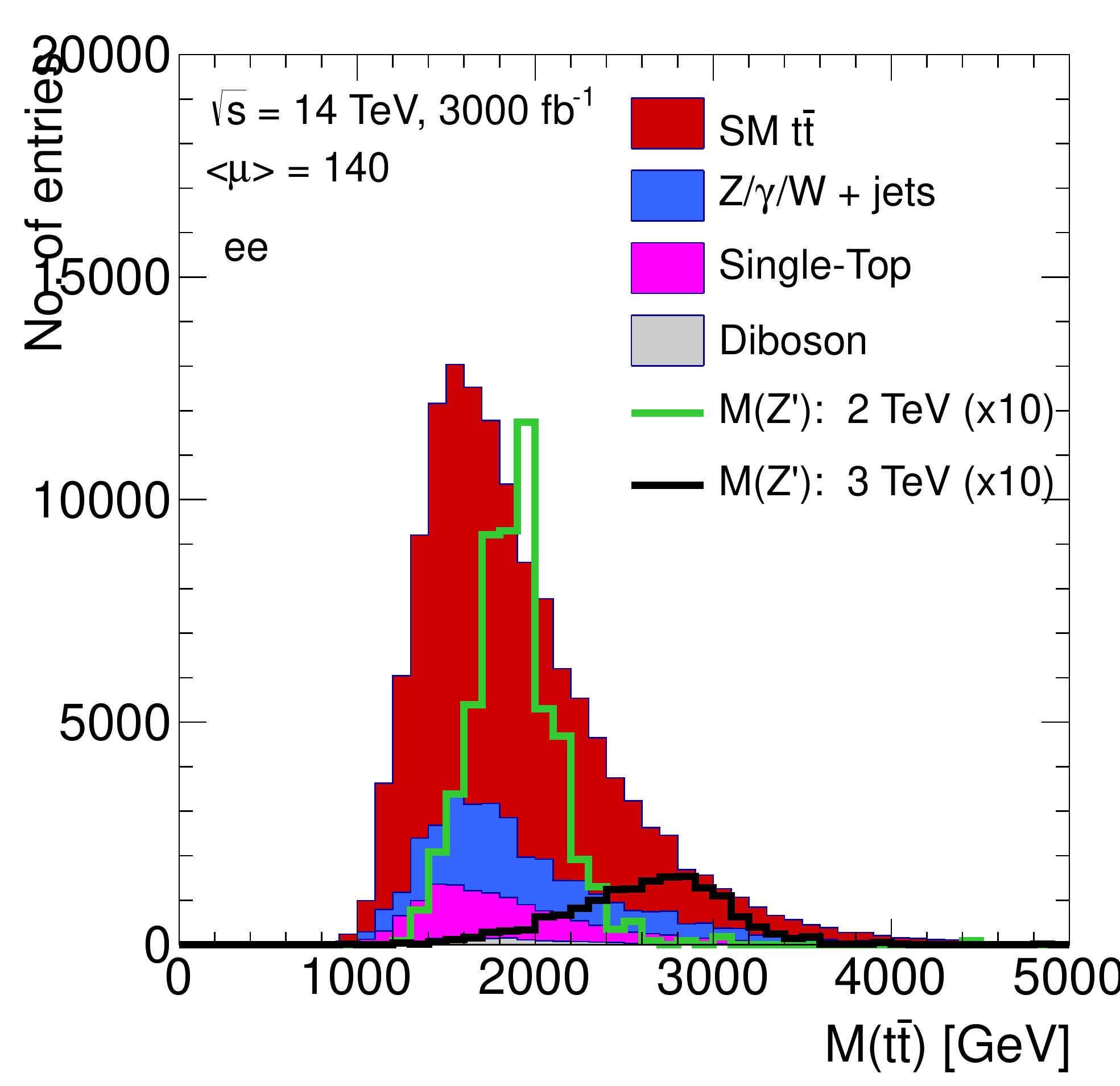}
  \includegraphics[width=0.33\textwidth]{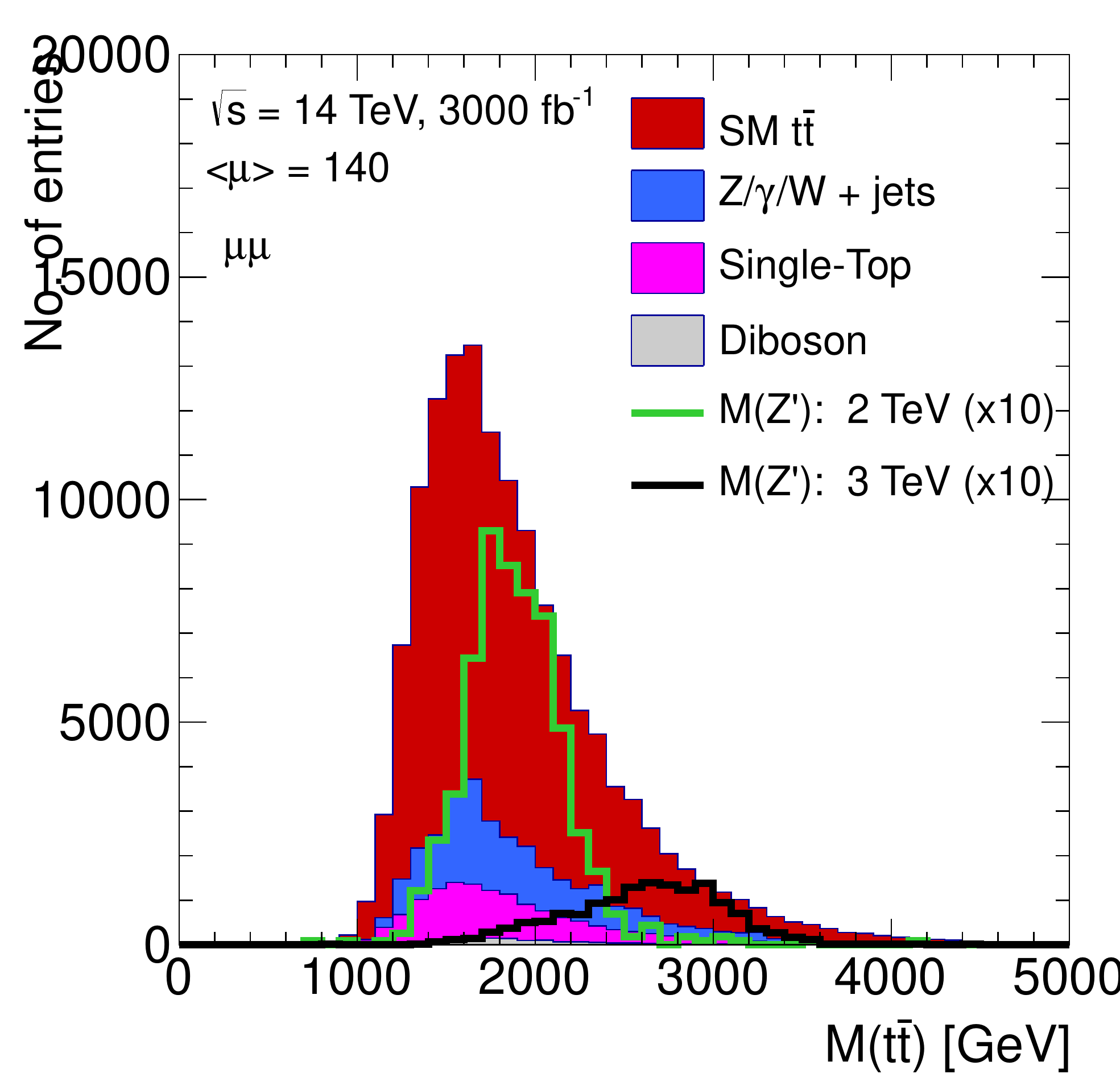}
  \includegraphics[width=0.33\textwidth]{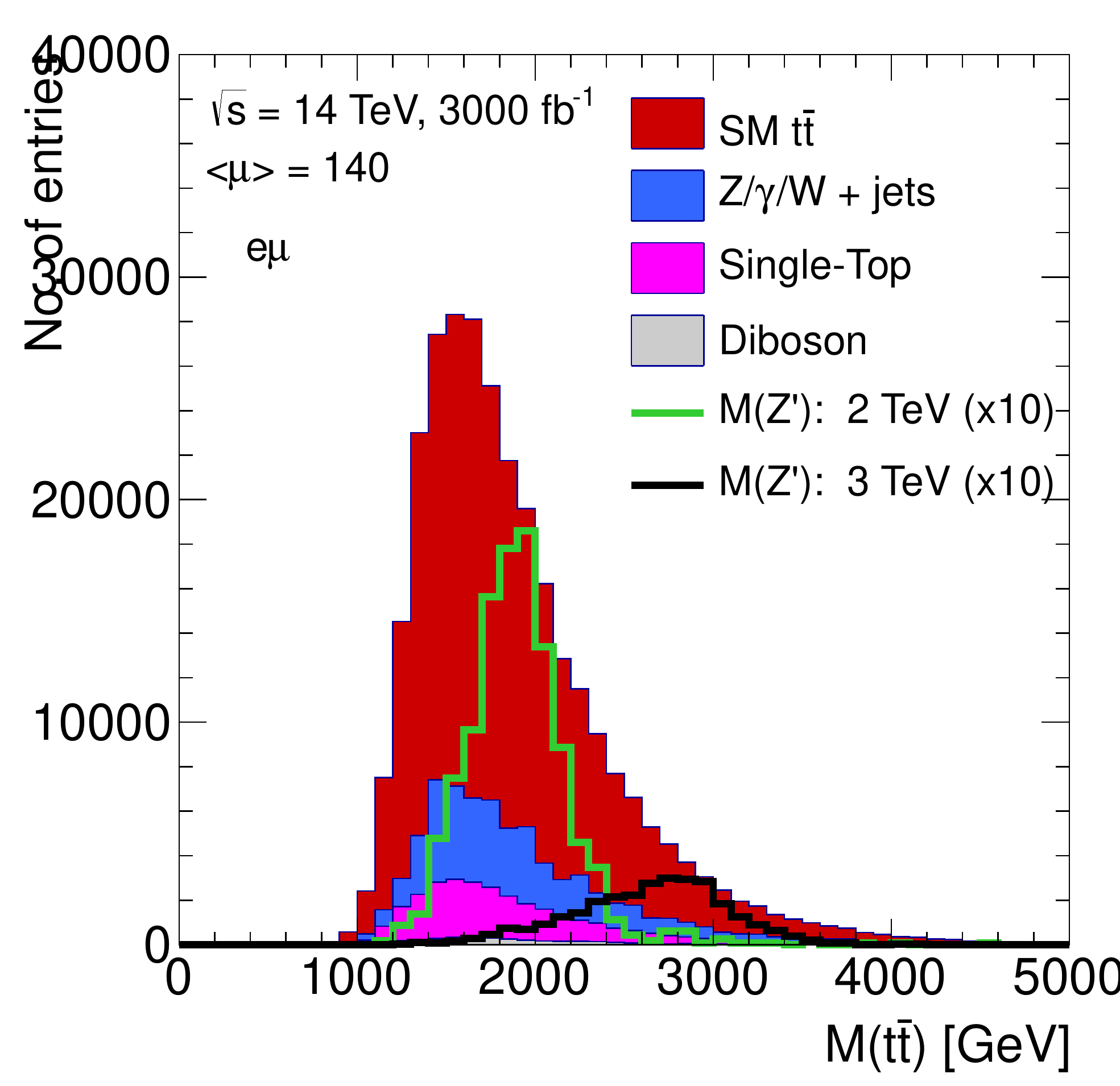}
\vspace{-0.12in}
 \caption{Distributions of the mass variable $M$ for the $ee$,
   $\mu\mu$, and $e\mu$ channels for 3000~$\rm{fb^{-1}}$ after selection cuts are applied.
\label{fig:invmass3000}}
\end{figure}

The other highly discriminating variable is  a mass variable $M$. 
The mass variable $M$ is computed
from the four-momenta of the two leading leptons, the two
leading jets and a four-momentum formed from the $p_x$ and $p_y$ 
components of the missing transverse momentum with the $p_z$ set to zero.
The distributions of the mass variable for the backgrounds and for the signal with
$\zprime$ masses of 2~TeV and 3~TeV, after \emph{all} selections, are
shown in Figs.~\ref{fig:invmass300} and~\ref{fig:invmass3000} for the
$\int\mathcal{L}\rm{dt} = 300$~$\rm{fb^{-1}}$ and 3000~$\rm{fb^{-1}}$ scenarios,
respectively. A heavy $\zprime$ produces higher values of $M$ than the background processes.
Table~\ref{tab:selections} summarizes the final selection cuts obtained
from the RGS for the two luminosity scenarios. The expected event yields are given 
in Table~\ref{tab:yields}.

%$\int\mathcal{L}\rm{dt} = 300$~(3000)~$\rm{fb^{-1}}$. 
%The resulting approximate $S/\sqrt{B}$ achieved is 4\%
%(15\%) for $\int\mathcal{L}\rm{dt} = 300$~(3000)~$\rm{fb^{-1}}$. 
%
%
%------------------------------------------------------------ 
% Table 1: Event selections  
%------------------------------------------------------------ 
\begin{table}[htbp]  
\begin{center}  
\caption{Summary of the final selection cuts obtained
from the RGS for the two LHC luminosity scenarios at $\roots=14$~TeV.}
\vspace*{0.2cm}  
\begin{tabular}{l|rr}\hline\hline 
LHC luminosity scenario  & \ \ \ \ \ $\int\mathcal{L}\rm{dt} = 300$~$\rm{fb^{-1}}$  
                         & \ \ \ \ \ $\int\mathcal{L}\rm{dt} = 3000$~$\rm{fb^{-1}}$ 
\\[3mm]
\hline
Leading lepton $p_T>$                     & 100 GeV  & 100 GeV \\ 
Second leading lepton $p_T>$  \ \ \       &  30 GeV  &  20 GeV \\ 
%Lepton $|\eta|<$ 2.4\\
Leading jet $p_T>$                        & 175 GeV  & 550 GeV \\ 
Second leading jet $p_T>$                 & 150 GeV  & 100 GeV \\ 
%Jet $|\eta|<$ 2.4\\
$\met >$                                  & 95 GeV   &  35 GeV \\ 
$\Delta R(\rm {lepton, closest~jet} ) < $ &  0.6  & 1.2 \\
$M >$                                     & 1500 GeV & -- \\
\hline\hline 
\end{tabular} 
\label{tab:selections} 
\end{center} 
\vspace*{-0.5cm} 
\end{table} 

\begin{table}[htbp]
\begin{center}
\caption{Summary of the expected signal and the background event
yields for the two LHC luminosity scenarios at $\roots=14$~TeV.}
\vspace*{0.2cm}
\begin{tabular}{|l|r|r|}\hline\hline
LHC luminosity scenario  & \ \ \ \ \ $\int\mathcal{L}\rm{dt} = 300$~$\rm{fb^{-1}}$
                         & \ \ \ \ \ $\int\mathcal{L}\rm{dt} = 3000$~$\rm{fb^{-1}}$
\\[3mm]
\hline
\multicolumn{3}{|c|}{Signal Event Yields}                    \\
\hline
$\zprime$ $\Mzprime$~=~2~TeV                      &  1395  &    22534   \\
$\zprime$ $\Mzprime$~=~3~TeV                      &   446  &     5955   \\
$\zprime$ $\Mzprime$~=~4~TeV                      &  85.7  &     1118   \\
$\zprime$ $\Mzprime$~=~5~TeV                      &  14.5  &      184   \\
\hline
\multicolumn{3}{|c|}{Background Event Yields}                \\
\hline
\ttbar                                            &  17599  &  427058   \\
single top                                        &   2044  &   50545   \\
$\rm{W}/\rm{Z}/\gamma^*$+jets                     &   2545  &   81740   \\
Diboson                                           &    163  &    6384   \\
\hline
Total background                                  &  22351  &  565727  \\
\hline\hline
\end{tabular}
\label{tab:yields}
\end{center}
\vspace*{-0.5cm}
\end{table}

\vspace*{3mm}

%----------------------------
% Results
%----------------------------
\section{Expected Discovery Reach and Limits }
\label{sec:results}

In order to quantify the expected 5$\sigma$ discovery or 95\% C.L. exclusion
limit for a $\zprime$ resonance, we use the Bayesian method~\cite{bayesian}
implemented in the statistical software
package {\sc theta}~\cite{theta}.
A multi-Poisson likelihood, constructed from the binned mass
distributions of all three channels ($ee$, $\mu\mu$, and
$e\mu$), is combined with a flat prior for the signal cross section. 
The following systematic uncertainties are
accounted for in the signal and background models, assuming full
correlation across channels: 10\% in the cross section
normalization for each background process, 10\% in the b-tagging
efficiency, and 2\% in the jet-energy scale.

\begin{figure}[hbtp]
\centering
 \includegraphics[width=0.48\textwidth]{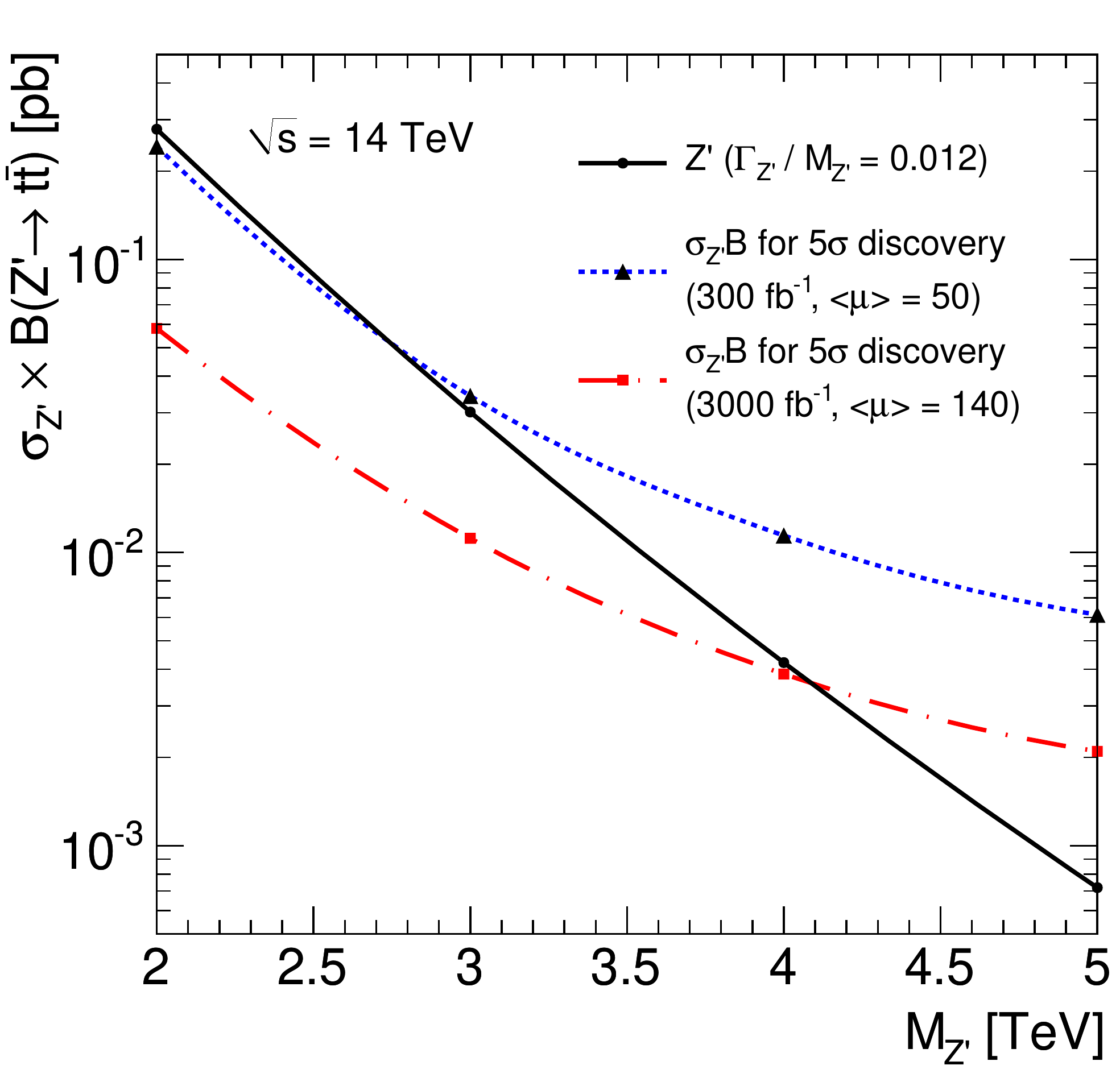}
 \includegraphics[width=0.48\textwidth]{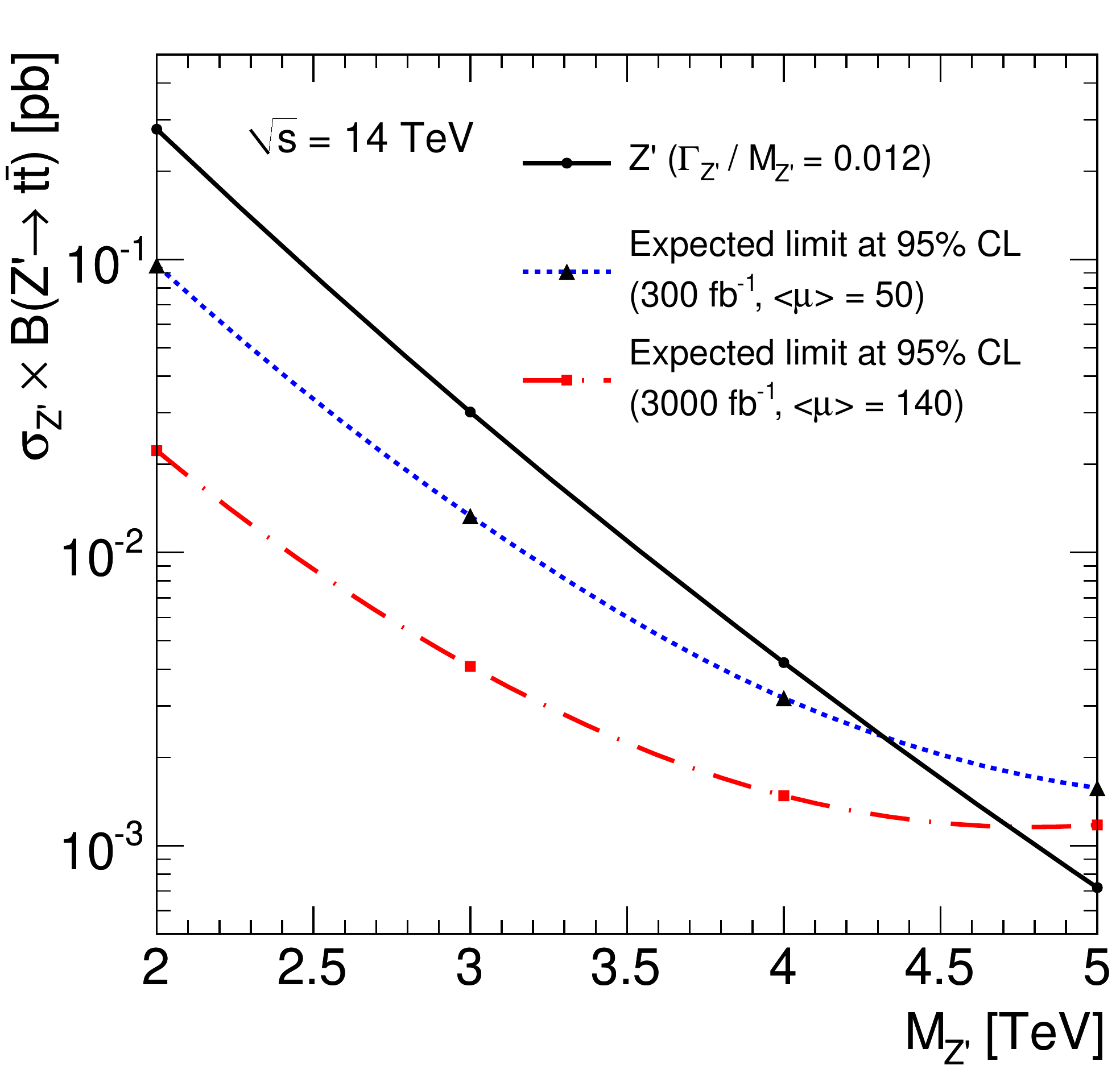}
\vspace{-0.12in}
\caption{Required $\sigma_{\zprime}\mathcal{B}$ for a 5$\sigma$
  observation (left) and upper limits at 95\% C.L. on 
  $\sigma_{\zprime}\mathcal{B}$ (right)  as a function of $\Mzprime$ 
for narrow-width, leptophobic $\zprime$ resonances. Also shown 
is the theoretical prediction for the $\zprime$.
\label{fig:limits}}
\end{figure}

Figure~\ref{fig:limits} (left) shows the $\zprime$ production cross section
times the branching fraction to $\ttbar$ ($\sigma_{\zprime}\mathcal{B}$), as a function of $\Mzprime$, that would yield a signal with a statistical significance of 5$\sigma$ at $\roots = 14$~TeV, that is, a discovery, with integrated luminosities
$\int\mathcal{L}\rm{dt} = 300~\rm{fb^{-1}}$ and $\int\mathcal{L}\rm{dt} = 3000~\rm{fb^{-1}}$. The  
cross section times branching fraction, $\sigma_{\zprime}\mathcal{B}$, ranges from
6 -- 300  (2 -- 60 )~fb with $\int\mathcal{L}\rm{dt} =  300~\rm{fb^{-1}}$ ($\int\mathcal{L}\rm{dt} = 3000~\rm{fb^{-1}}$)
for the mass range 2--5~TeV.
Comparing these with the theoretical prediction for  the production cross section of 
a leptophobic $\zprime$ yields the expected $\zprime$ discovery mass reach
of 2.8~TeV with $\int\mathcal{L}\rm{dt} = 300~\rm{fb^{-1}}$ and 4.1~TeV with $\int\mathcal{L}\rm{dt} = 3000~\rm{fb^{-1}}$.  

Figure~\ref{fig:limits} (right) shows expected 95\% C.L. limits on $\sigma_{\zprime}\mathcal{B}$
as a function of $\Mzprime$ for the two luminosity scenarios. The 
expected limits range from 2 -- 100~(1--20)~fb with $\int\mathcal{L}\rm{dt} = 300~\rm{fb^{-1}}$ ($\int\mathcal{L}\rm{dt} = 3000~\rm{fb^{-1}}$)
for the mass range 2--5~TeV. 
Comparing these with the predicted production cross section for a leptophobic $\zprime$
shows that we can expect to exclude 
the existence of a $\zprime$ with mass
$< 4.4$~(4.7)~TeV at 95\% C.L. with $\int\mathcal{L}\rm{dt} = 300~\rm{fb^{-1}}$ ($\int\mathcal{L}\rm{dt} = 3000~\rm{fb^{-1}}$) 
should we fail to make a discovery.

%

%----------------------------
% Summary
%----------------------------
\section{Summary}
\label{sec:summary}

We have  assessed the potential for finding 
evidence of a leptophobic $\zprime$ boson in
$\zprime \rightarrow \ttbar \rightarrow 2\ell+2\nu+\bbbar$
decays in $\pp$
collisions at $\roots$~=~14~TeV. Two sets of hypothetical 
data, simulated using {\sc pythia}, {\sc Madgraph}  and {\sc Delphes},
have been analyzed assuming an integrated luminosity of 
$\int\mathcal{L}\rm{dt} = 300~\rm{fb^{-1}}$ with an average number of 
events per bunch crossing (pile-up) of $\rm{<PU>}=$~50, and 
$\int\mathcal{L}\rm{dt} = 3000~\rm{fb^{-1}}$ with $\rm{<PU>}=$~140. 
For the lower (higher) integrated luminosity, our study indicates that it
is possible to discover a $\zprime$ up to a mass 2.8~(4.1)~TeV with a statistical
significance of 5$\sigma$. Should we fail to make a discovery, 
the existence
of a $\zprime$ with mass $< 4.4$~(4.7)~TeV can be excluded at 95\%
C.L. using data associated with the lower (higher) integrated
luminosity scenario. 

%----------------------------
% Acknowledgments
%----------------------------
\section*{Acknowledgments}
\label{sec:acknowledge}

The authors would like to thank James Dolen, John Stupak, Sergei
Chekanov, and Johannes Erdmann for their help in setting up tools 
as well as providing the samples for this study. 
%James Dolen -- provided all initial reference material for startup
%John Stupak -- helped set up the batch tool on condor
%Sergei Chekanov -- provided details of official MC samples
%Johannes Erdmann -- provided the $\zprime$ samples at additional mass points

%%%%%%%%%%%%%%%%%%%%%%%%%%%%%%%%%%%%%%%%%%%%%%%%%%%
% Bibliography
%%%%%%%%%%%%%%%%%%%%%%%%%%%%%%%%%%%%%%%%%%%%%%%%%%%
%\clearpage

\end{document}